# Enhancing understanding and clinical applications of cerebral autoregulation: A novel integrated numerical framework

Qi Zhang[a,b,d]; Meng-di Yang[c]; Xuan-hao Xu[b]; Xiu-li Xu[e]; Shuai Tian[b,*]; Li-ling Hao[a,*]

a   *College of Medicine and Biological Information Engineering, Northeastern University, Shenyang, Liaoning, 110167, China.*

b   *Department of Cardiology, The Eighth Affiliated Hospital of Sun Yat-sen University, Shenzhen, Guangdong. 518033, China.*

c   *Department of Critical Care Medicine, Beijing Haidian Hospital, Beijing, 100080, China.*

d   *The School of Biomedical Engineering, Sun Yat-sen University, Shenzhen, Guangdong, 518107, China.*

e   *Department of Emergency, The Eighth Affiliated Hospital of Sun Yat-sen University, Shenzhen, Guangdong. 518033, China.*

## Abstract

**Background:** Cerebral autoregulation (CA) is a fundamental mechanism that modulates cerebrovascular resistance, primarily by regulating the diameter of small cerebral vessels to maintain stable cerebral blood flow (CBF) in response to fluctuations in systemic arterial pressure. However, the clinical understanding of CA remains limited due to the intricate structure of the cerebral vasculature and the challenges in accurately quantifying the hemodynamic and physiological parameters that govern this autoregulatory process. **Method:** In this study, we introduced a novel numerical algorithm that employs three partial differential equations and one ordinary differential equation to capture both the spatial and temporal distributions of key CA-driving factors, including the arterial pressure ($P$) and the partial pressures of oxygen ($PO_2$) and carbon dioxide ($PCO_2$) within the cerebral vasculature, together with a Windkessel model in turn to regulate the CBF based on the calculated $P$, $PO_2$, and $PCO_2$. This algorithm was sequentially integrated with our previously developed personalized 0D-1D multi-dimensional model to account for the patient-specific effects. **Results:** The integrated framework was rigorously validated using two independent datasets, demonstrating its high reliability and accuracy in capturing the regulatory effects of CA on CBF across a range of physiological conditions. **Conclusion:** This work significantly advances our understanding of CA and provides a promising foundation for developing hemodynamic-based therapeutic strategies aimed at improving clinical outcomes in patients with cerebrovascular disorders.

* Corresponding Authors
  *E-mail addresses*: tiansh9@mail.sysu.edu.cn (Shuai Tian); haoll@bmie.neu.edu.cn (Li-ling Hao)

## 1. Introduction

Cerebral blood flow (CBF) is essential for maintaining the brain's physiological functions, supplying oxygen and nutrients critical for neural activity [1][2]. The regulation of CBF is primarily governed by cerebral autoregulation (CA), a critical mechanism that ensures the brain receives an adequate and consistent blood supply despite fluctuations in physiological conditions (for example, systemic arterial pressure) [3][4]. Given its pivotal role, CA is of great importance in the management of cerebral disorders, particularly in conditions such as strokes and traumatic brain injury, making it a key focus of clinical research [5].

CA can be broadly categorized into two types, i.e., static and dynamic autoregulation, based on the temporal scale over which they operate [3], [4]. Static autoregulation refers to the long-term regulatory mechanisms that maintain CBF to meet sustained physiological needs. In contrast, dynamic autoregulation operates on shorter periods, responding rapidly to acute, transient changes in blood pressure or other physiological disturbances. Although these two types of autoregulation operate over different time scales, recent studies have shown that they often produce consistent results, suggesting a degree of overlap in their mechanisms [5]. Both types of autoregulation depend on vascular reactivity, influence each other during their respective operations, and ultimately work together to maintain stable CBF. Due to its feasibility in measurement, dynamic autoregulation plays a particularly crucial role in advancing our understanding of the underlying mechanism of CA [5].

Multiple hypotheses have been proposed to explain the regulation of CA, including metabolic regulation [6], myogenic regulation [7], neurogenic regulation [8], and endothelial regulation [9]. Among these, metabolic regulation, particularly in relation to the partial pressures of oxygen ($P_{O_2}$) and carbon dioxide ($P_{CO_2}$) in the cerebral circulation, is widely regarded as the most influential. $CO_2$, in particular, functions as a potent vasodilator, driving an increase in CBF in response to elevated $CO_2$ or reduced $O_2$ levels [6], [10]. This vasodilation is particularly critical in the rapid regulation of CBF, ensuring adequate oxygen and nutrient delivery to the brain during periods of fluctuating metabolic demand.

However, drawing explicit conclusions from clinical experiments has proven to be challenging, as CA is not a straightforward mechanism. It involves multiple interacting factors and processes, which operate interactively rather than independently to form a multifaceted response [10]. Consequently, while many hypotheses have been proposed to explain different clinical phenomena, no single mechanism has yet been able to account for all observed clinical outcomes. Numerous clinical trials have even failed to provide an accurate empirical formula that can be directly applied in clinical practice. This highlights the need for innovative and reliable methods to better understand the underlying mechanism of CA.

Computational Fluid Dynamics (CFD) has emerged as a pivotal tool for demonstrating the mechanism of CA, offering significant advantages over traditional clinical approaches due to its low cost and risk-free nature [11], [12]. The pioneering work by Panerai et al. employed an integrated approach that coupled empirical formulas with reduced-order models to estimate CBF under CA [13]. However, these empirical formulas were limited in overlooking both spatial and temporal variation, unavoidably affecting the accuracy of the predictions. Subsequent works utilized ordinary differential equation (ODE) to describe boundary conditions, thereby improving the accuracy of CA-driving factors (blood pressure, $PO_2$, $PCO_2$) by incorporating temporal variations [14]-[16]. However, the distribution of CA-driving factors in the vascular tree, particularly with respect to patient-specific variations, remains unclear, limiting broader application in real clinical practice.

(a) Clinical task: The CBF prediction under CA

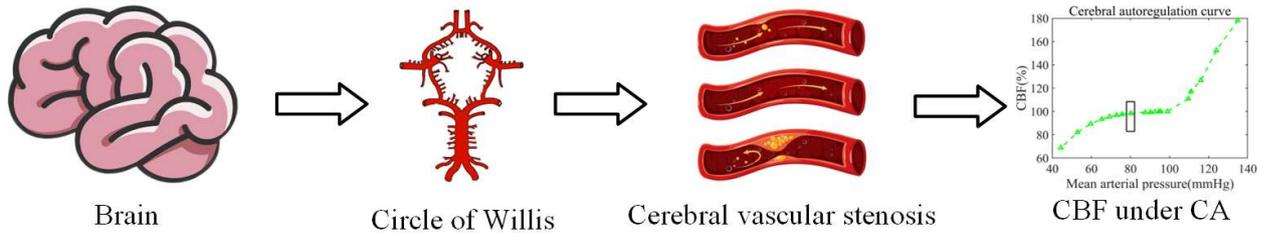

(b) Challenge: Complex process driven by multiple factors

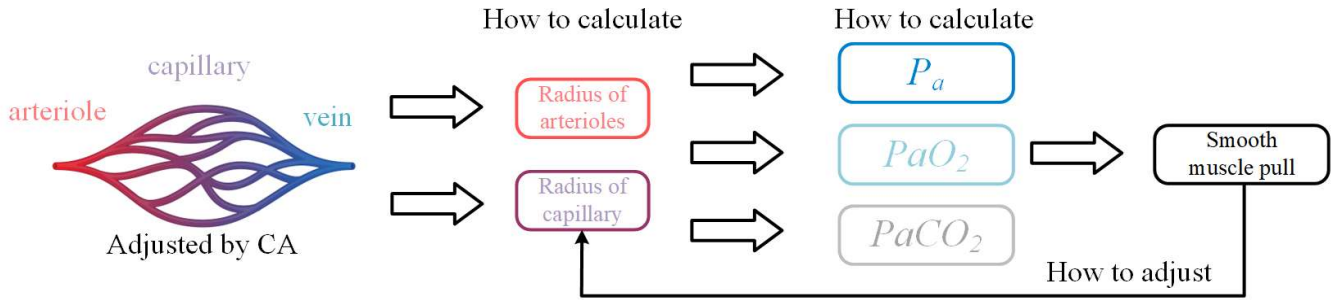

(c) Our solution: PDEs-ODE model

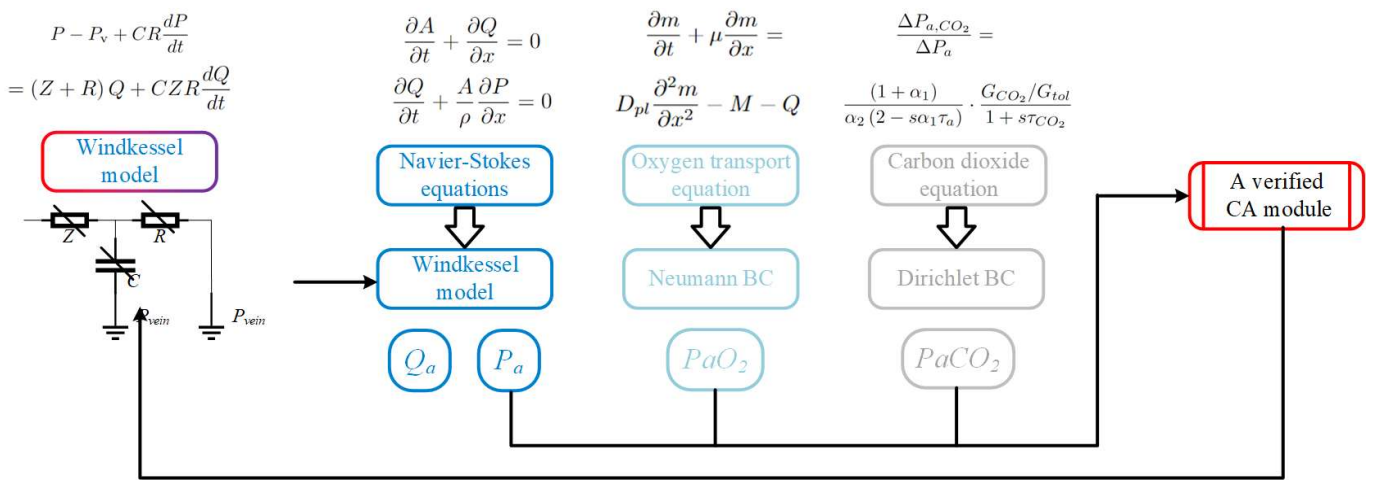

Figure 1. The flow of this research (a) Clinical task: calculate the CBF under CA. (b) Challenge: Complex process driven by multiple factors. (c) Our solution: PDEs-ODE model.

To address these challenges, the present study introduced a novel numerical model to evaluate both the temporal and spatial distribution of key CA-driving factors [17], [18]. Specifically, as shown in Figure 1, one set of partial differential equations (PDEs) were used to predict arterial pressure ($P$), a PDE was employed to determine $PO_2$ [19][20], an ODE was utilized to evaluate $PCO_2$ [20], and a Windkessel model, which could be adjusted according to the solved hemodynamic factors, was developed for the boundary condition specification [21]. This model was used to replace the cerebral circulation compartment of our previously developed 0D-1D model which could account for the personalized characteristics, ultimately resulting in a comprehensive numerical framework for investigating CA that incorporated both physiological dynamics and patient-specific variations. This integrated framework was rigorously validated for its great accuracy and reliability across two datasets, virtual dataset contains healthy, stenosis with CA, and stenosis without CA and clinical dataset with CBF data from ischemic stroke patients in different scenarios. This work advances our understanding of CA and provides a promising foundation for developing hemodynamic-based therapeutic strategies aimed at improving clinical outcomes in patients with cerebrovascular disorders.

## 2. Methods

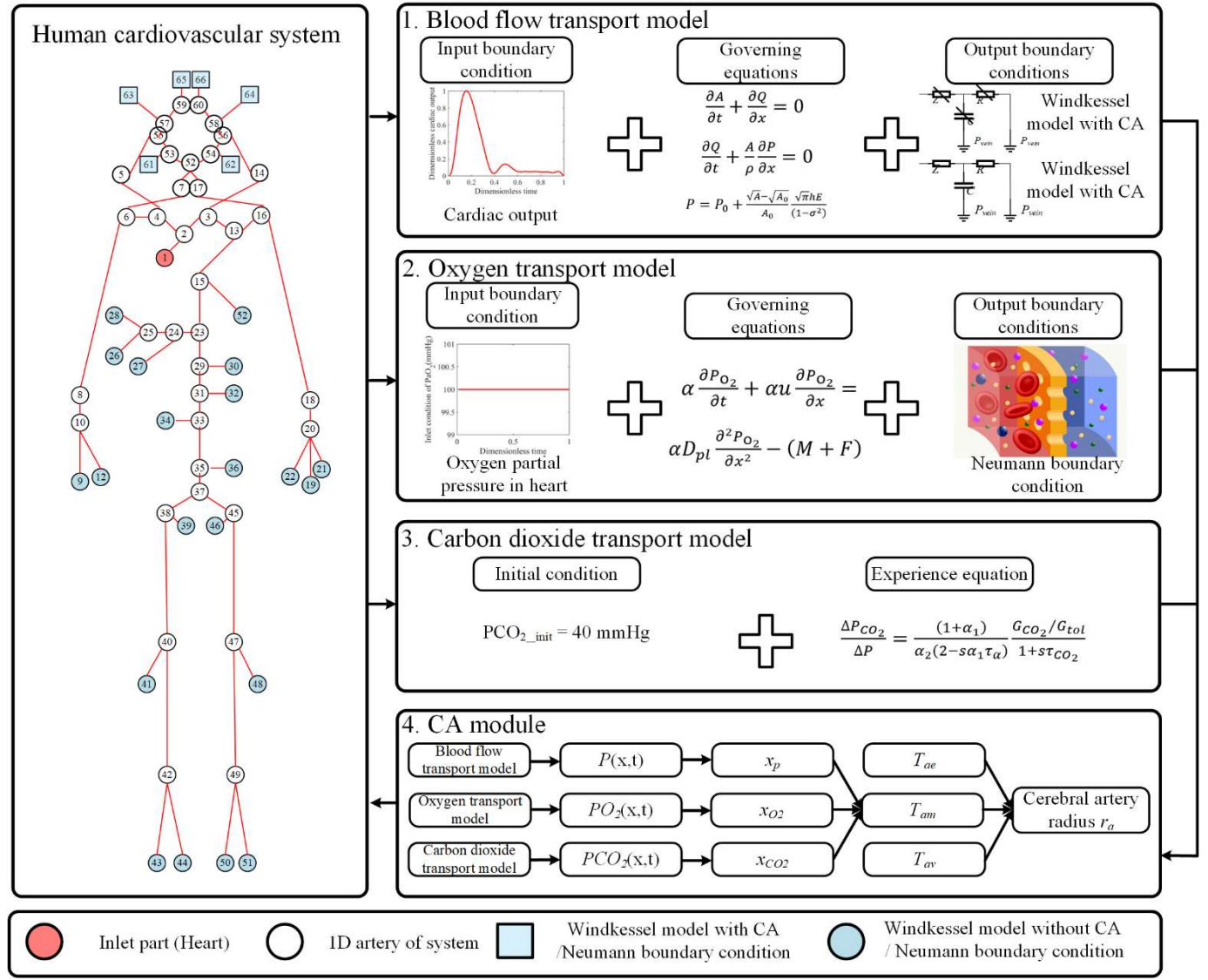

Figure 2. The methodology of this research.

2.1. Blood flow transport model.

The cerebral arterial network is characterized by intricate curvatures and complex interconnections, making it challenging to capture the 3D distribution of hemodynamics. These geometric complexities lead to significant computational demands and extended processing times. To address this, we employed a reduced-order 1D model, which has demonstrated its high reliability and accuracy in describing blood flow behavior within a vascular tree, particularly along the axial direction in our previous work [11][12], and thus, (illustrated in Figure 2) inheritably adopted in this study. This model comprises 74 segments of 1D tubes that represent the arterial network and 21 sets of Windkessel models to simulate the arterioles and capillaries. The blood rheology is assumed to follow Newtonian fluid behavior ($\mu$ = 0.003 Pa·s), and the analysis is restricted to unidirectional flow patterns.

The mass and momentum conservation equations governing this 1D model are as follows:

$$\frac{\partial A}{\partial t} + \frac{\partial Q}{\partial x} = 0 \tag{1}$$

$$\frac{\partial Q}{\partial t} + \frac{A}{\rho}\frac{\partial P}{\partial x} = 0 \qquad (2)$$

where $t$ is the time, $x$ is the longitudinal coordinate along the vessel, $Q$ is the blood flow rate, $A$ is the vessel cross-sectional area, and $\rho$ is the blood density (assumed to be 1060 kg/m³).

Fluid-structure interaction was considered at the vessel wall to account for vessel deformation [22]. Vessel compliance is modeled by allowing the cross-sectional area to change, as described by the following equation coupled with Eqs. (1) and (2):

$$P = P_0 + \frac{\sqrt{A}-\sqrt{A_0}}{A_0}\frac{\sqrt{\pi}hE}{(1-\sigma^2)} \qquad (3)$$

where $P_0$ is the reference pressure, $A_0$ and $h$ are the cross-sectional area and wall thickness of the vessel at the reference pressure, $E$ is the Young's modulus, and $\sigma$ is the Poisson's ratio (= 0.45).

2.2. Oxygen transport model.

In this study, a proportional relationship was assumed between the concentration of dissolved oxygen and its partial pressure (according to Henry's law). Therefore, the transport equation for the dissolved oxygen concentration in the blood is given by [23],

$$\alpha\frac{\partial PO_2}{\partial t} + \alpha u\frac{\partial PO_2}{\partial x} = \alpha D_{pl}\frac{\partial^2 PO_2}{\partial x^2} - (M + F) \qquad (4)$$

where $\alpha$ is the oxygen dissolution coefficient in plasma (= 2.5×10⁻⁵ mol/(L\Pa)), $D_{pl}$ is the diffusion coefficient of free oxygen in blood (= 1.2×10⁻⁵ m²/s [23]), and $(M+F)$ represents the oxygen consumption, with M indicating a constant volumetric consumption rate of oxygen by cells within the arterial wall tissue (= 2.1×10⁵ L/s [23]) and $F$ representing the mass of oxygen crossing the surface of the artery wall, which could be given as follow [21], [24],

$$F = \frac{2}{r}D_w\frac{\alpha PO_2 - \alpha_t PO_2 t}{h} \qquad (5)$$

where $r$ is the vessel radius, $D_w$ defines the diffusion coefficient of oxygen in the arterial wall, $\alpha_t$ is the coefficient of solubility in the tissue, and $PO_2 t$ is the oxygen tension of the tissue (= 40 mmHg).

2.3. Carbon dioxide transport model.

The relationship between arterial pressure and $CO_2$ concentration is critical in regulating CBF [21]. In this investigation, an empirical formula obtained from previous studies was used to co-relate $PCO_2$ and $P$ in the end of the cerebral artery as follow:

$$\frac{\Delta PCO_2}{\Delta P} = \frac{(1+\alpha_1)}{\alpha_2(2-s\alpha_1\tau_\alpha)}\frac{G_{CO_2}/G_{tol}}{1+s\tau_{CO_2}} \qquad (6)$$

where $\alpha_1$ is the time constant (= 0.561), $\alpha_2$ represents the nonlinear feedback gain of CBF (= 0.286), $\tau_\alpha$ is the arterial inflow time constant (= 1.24 s), $G_{CO_2}/G_{tol}$ represents the ratio of the feedback for $CO_2$, with a value of 1/3 (three factors co-effects), $\tau_{CO_2}$ defines the feedback time constant of $CO_2$ with value of 40s. Based on Eq. (11), $PCO_2$ is calculated to drive CA.

2.4. Module of CA.

As stated above, CA is achieved through changing the vessel diameter by regulating the cerebral vascular smooth muscle. In this study, three activators, $x_P$, $x_{O_2}$, and $x_{CO_2}$, were worked out through the calculated $P$ $PO_2$, and $PCO_2$, respectively. These activators were used to account for their effects on regulating the muscular tension for fine-tuning the internal vessel diameter in anterior cerebral artery (ACA), middle cerebral artery (MCA), and posterior cerebral artery (PCA) [14],[18],[19]. A co-activation factor could be defined as follow,

$$M_s = \frac{e^{M_{s1}} - 1}{e^{M_{s1}} + 1} \tag{7}$$

where

$$M_{s1} = M_P \cdot x_P + M_{O_2} \cdot x_{O_2} + M_{CO_2} \cdot x_{CO_2} \tag{8}$$

here, $M_P$, $M_{O_2}$, and $M_{CO_2}$ represent the weight factors with values of 0.034, 1.00, and 2.00, corresponding to $P$, $PO_2$, and $PCO_2$, respectively. Then, the muscular tension $T_{am}$ could be given by,

$$T_{am} = T_{am0} \cdot (1 + M_s) \cdot e^{-\left|\frac{r_a - r_{a0}}{r_{a0}}\right|^{n_{am}}} \tag{9}$$

where $T_{am0}$ is the maximum smooth muscle tension in the basal state and assumed to be 1.5 mmHg·s in this study, $r_{a0}$ and $r_a$ represent the initial distal arterioles radius and actual distal arterioles radius, respectively, $n_{am}$ is the constant parameter of the smooth muscle tension model with a value of 1.75. In addition, two other tensions, elastic tension ($T_{ae}$) and viscous tension ($T_{av}$) are considered in this study. Their definition are as follows,

$$T_{ae} = \left|\sigma_{ae0} \cdot \left(e^{K_{ae}\frac{r_a - r_{a0}}{r_{a0}}} - 1\right)\right| \cdot h_a \tag{10}$$

$$T_{av} = \frac{\eta_a}{r_{a0}} \cdot \frac{r_a - r_{a0}}{T_{cyc}} \cdot h_a \tag{11}$$

Thus, the total tension $T_a$ is,

$$T_a = T_{ae} + T_{am} + T_{av} \tag{12}$$

where $\sigma_{ae0}$, $K_{ae}$, and $\sigma_{ac}$ represent the constant parameters of the elastic stress model with values of 11.19, 4.5 and 41.32 mmHg, respectively. The wall thickness of arterioles $h_a$ can be worked out as,

$$h_a = \sqrt{r_a^2 + 2r_{a0}h_{a0} + h_{a0}^2} - r_a \tag{13}$$

where, $h_{a0}$ represents the initial thickness of the arterioles. The relationship between the radius of arterioles and different tensions is established through Eqs. (12)-(13), but the $r_a$ is still unknown in this stage. To elucidate the alterations in arterioles influenced by CA, the pressure-tension relationship is incorporated and simultaneous to Eq. (12) to dynamically derive $r_a$.

$$T_a = P_{ma}r_a - P_{ic}(r_a + h_a) \tag{14}$$

where $P_{ma}$ is the mean arteriole pressure, derived from the iterative solution of the previous cardiac cycle [12], and $P_{ic}$ is the intracranial pressure, assumed to be a constant value of 10 mmHg.

2.5. Simulations

A system of equation incorporating these four components was constructed to describe the hemodynamic response, following physiological principles and clinical observations. In the discretization process, the time interval (i.e., Δt) was set to be 1/80 of the cardiac cycle (~0.00625 - 0.015 s in this study), and the spatial interval was set to 0.005 m. At the inlet boundary, the flow rate, $Q_{in}(t)$ was determined based on clinical data, which means the volume per beat is based on the patient's weight [12]. A three-element Windkessel model was applied at the outlets [23]. Its variables, the arterial resistance (Z), arterial compliance (C), and arteriole/capillary resistance (R), are determined as follows [12],

$$Z + R = \frac{\sum_{j=1}^{n} A_j}{A_j} \cdot \frac{MAP}{Q_{ao}}, \qquad (15)$$

$$C = \frac{A_j}{\sum_{j=1}^{n} A_j} C_{tot} \qquad (16)$$

$$Z = 0.2R \qquad (17)$$

where, $A_j = \pi r_j^2$, $C_{tot}$ = 8.93 × $10^{-12}$ $m^5$/N, and $j$ represents the index of vessel connecting to the Windkessel model. Thus, the arterial pressure $P$ could be obtained as follow,

$$P - P_v + CR\frac{dP}{dt} = (Z + R)Q + CZR\frac{dQ}{dt} \qquad (18)$$

where, $P_v$ is the venous pressure, assumed to be 10 mmHg in this study.

The overall flow domain was initialized with a mean arterial pressure (MAP) of 100 mmHg and a flow rate of 1 mL/s, except $Q_{in}$ which was adjusted by patient's weights. The Eqs. (1)-(3) were solved using an implicit finite difference scheme for pressure and flow [24]. Specifically, the upwind scheme was used to discretize the Eqs. (4)-(5), and nonlinear terms were solved using the Newton-Raphson method. The Navier-Stokes equation and boundary conditions were discretized and parameterized by clinical measurements and others estimations [12], and the final form was shown in the follow,

$$K \cdot P_t = f + Q_t \qquad (19)$$

where the $K$ is determined according to the geometric characteristics [26], and the subscript $t$ represents the time point. Final solutions were assumed when the maximum relative difference for pressure, flowrate, $PCO_2$, and $PO_2$ between two consecutive cardiac cycles reached 1%.

## 3. Experiments and Results

To verify the accuracy of the proposed method, two datasets of one virtual dataset and one clinical dataset were utilized.

### 3.1. Validation in the virtual dataset

We first verify the stability and validity of the model on the virtual dataset.

#### 3.1.1. The building of virtual dataset

Stenosis is defined as the abnormal narrowing of a vessel and resulting in increased vessel resistance, which poses a significant risk factor for cerebral ischemic stroke. The resultant reduction in distal cerebral blood flow may activate CA mechanisms. Validating the role of CA in various stenotic lesions requires substantial clinical data. However, obtaining such data from clinical trials is challenging due to the complexity of CA and the variability in patient responses. To address this, virtual datasets was generated through building multiple stenosis. Specifically, we focused on two common types of stenosis that are clinically relevant and have significant impact on cerebral hemodynamics: intracranial stenosis (such as middle cerebral artery MCA, stenosis) and extracranial stenosis (such as internal carotid artery, ICA stenosis). These models were chosen because they represent the most frequent and clinically significant sites of stenosis that can lead to cerebral ischemia.

Considering MCA stenosis as the most prevalent intracranial artery stenosis in clinical settings, with significant implications for stroke treatment, different degrees of MCA stenosis models were integrated into our proposed model. As the primary blood supply artery for the ACA and MCA, the carotid artery exerts a significant influence on cerebral hemodynamic changes.

ICA stenosis, a common extra-cranial stenosis, poses a higher risk of precipitating extensive ischemia when collateral circulation is compromised. To investigate this disease, we modeled severe stenosis (90%) of the right internal carotid artery and integrated the CA model to simulate blood flow changes post-regulation.

The degree of stenosis was quantified as follow: degree of stenosis = 1 - (minimum area of vessel in stenosis) / (unobstructed vessels area) * 100%. The formula provides a standardized method to assess the severity of stenosis, which is crucial for understanding its impact on cerebral blood flow. Regarding the effect of stenosis on hemodynamics, Young et al initially proposed a formula for the pressure drop across stenosis, which has been further refined in subsequent studies, as shown in Eq. (20) [27].

$$\Delta p = K_v \frac{\mu}{2\sqrt{A_s/\pi}} U + \frac{K_t}{2} \frac{(A_o - A_s)^2}{A_s^2} \rho |U| U + K_u \rho L \frac{dU}{dt} \tag{20}$$

where the U represents the velocity of flow, and Ao represents the original cross-sectional area of the stenosis vessel, As is the minimum cross-sectional area at stenosis, L is the length of stenosis, $K_l$ and $K_u$ are the fixed parameters with the values of 1.5 and 1.2, and $K_v$ can be written as follow format.

$$K_v = 32 \left( 0.83L + 1.64\left(2\sqrt{A_s/\pi}\right) \right) \cdot \frac{A_o^2}{A_s^2 \sqrt{A_s/\pi}} \tag{21}$$

Then, the Eqs. (20) and (21) were coupled in Eq. (2) as a force source term for momentum conservation [25], enabling the investigation of the effects of intracranial and extra-cranial stenosis.

3.1.2. The results of PDEs-ODE model on virtual dataset

To evaluate the effects of PDEs-ODE model, we input a virtual dataset representing three states—healthy, stenosis without CA, and stenosis with CA—into our model.

**CA model for MCA stenosis**. We investigated the impact of severe MCA stenosis (90%) on CBF and the role of the CA in maintaining blood flow stability. Compared to the healthy state, severe MCA stenosis without CA significantly reduced blood flow by approximately 18% and mean arterial pressure by 17 mmHg. This highlights the substantial impact of severe stenosis on cerebral hemodynamics in the absence of compensatory mechanisms... As shown in Fig. 3, the proposed CA model effectively maintains a constant blood flow despite a decrease in MCA mean arterial pressure from 100 mmHg to 92 mmHg. This demonstrates the critical role of CA in stabilizing CBF under conditions of severe stenosis. However, it is important to note that while CA model enhances smooth muscle tone associated with autoregulatory capacity, the overall CA ability is still weakened due to the decreased arterial pressure in the MCA and its distal arterioles. Additionally, we observed that the pressure in the MCA decreased due to the stenosis and further declined even with the presence of CA. This suggests that while CA helps to maintain blood flow, it cannot completely counteract the pressure drop caused by severe stenosis, aligning with previous clinical findings.

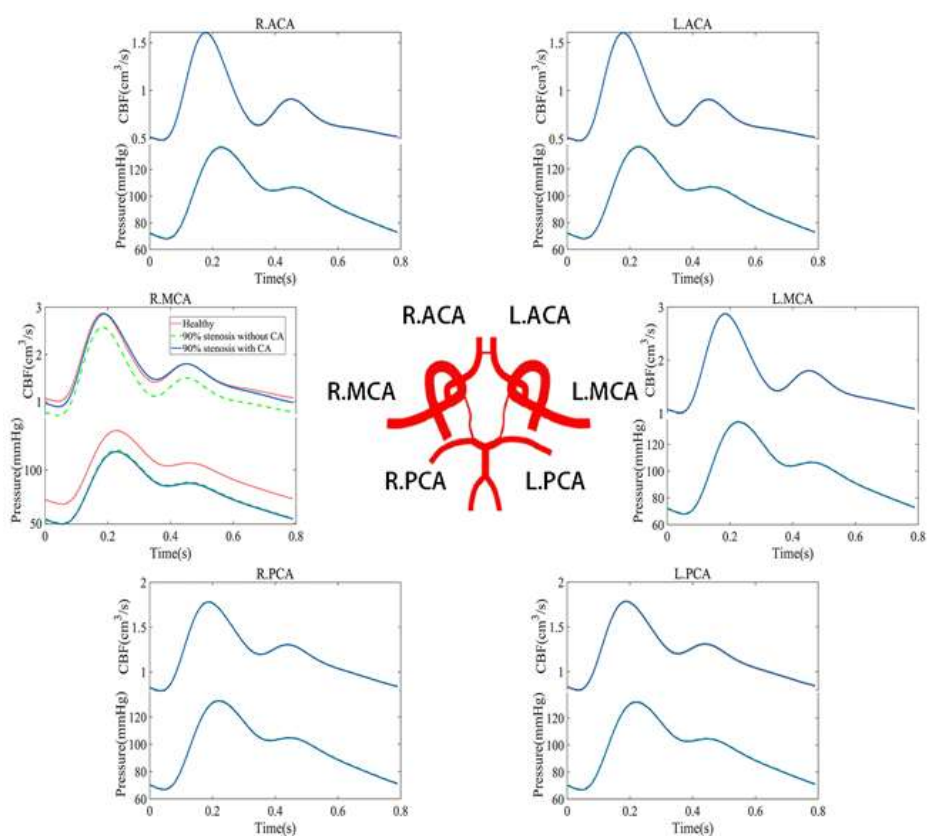

Figure 3. The flow and pressure waveforms of CoW with severe (90%) stenosis of the right MCA (R.MCA). Red line represents virtual subject without stenosis, green line represents 90% stenosis in the MCA with CA, and blue line represents 90% stenosis in the MCA without CA.

**CA model for ICA stenosis.** The ICA stenosis leads to inadequate blood supply to all cerebral vessels on the affected side. In this study, we utilized a complete Circle of Willis (CoW) model to simulate the blood supply to the ACA and MCA on the lesion side. This model ensured that the ACA and MCA received blood supply from the ipsilateral carotid artery to the anterior communicating artery, resulting in alterations in patient $PO_2$ and subsequent outcomes, as depicted in Table 1. The model of R.ICA stenosis revealed that the R.ACA and R.MCA exhibit more than 10% ischemia in the absence of CA, with evident effects extending to the left brain, resulting in decreased blood flow. However, bilateral PCA exhibit minimal reduction, and blood flow in the right brain slightly lags behind that in the left brain, primarily due to the influence of the R.PCA. Upon introducing the CA model, alterations in the CBF supply pathway are observed, with the R.ACA and R.MCA primarily supplied through the left internal carotid artery and anterior communicating artery.

Table 1: The mean flow, mean blood pressure, and arteriolar radius of cerebral arteries with severe stenosis (90%) of the R.ICA.

| Arteries | Variables | Health case | R.ICA stenosis without CA | R.ICA stenosis with CA |
|---|---|---|---|---|
| R.ACA | Q (mL/s) | 0.81 | 0.68 ↓ | **0.80** |
| | P (mmHg) | 98.27 | 82.56 ↓ | 81.20 ↓ |
| | $r_a$ (cm) | 0.0061 | 0.0061 | 0.0064 ↑ |
| L.ACA | Q (mL/s) | 0.81 | 0.72 ↓ | **0.79** |
| | P (mmHg) | 98.23 | 87.46 ↓ | 86.26 ↓ |
| | $r_a$ (cm) | 0.0061 | 0.0061 | 0.0062 ↑ |
| R.MCA | Q (mL/s) | 1.62 | 1.30 ↓ | **1.56** |
| | P (mmHg) | 98.21 | 79.34 ↓ | 77.90 ↓ |
| | $r_a$ (cm) | 0.0065 | 0.0065 | 0.0069 ↑ |
| L.MCA | Q (mL/s) | 1.62 | 1.49 ↓ | **1.58** |
| | P (mmHg) | 98.11 | 90.53 ↓ | 89.44 ↓ |
| | $r_a$ (cm) | 0.0065 | 0.0065 | 0.0066 ↑ |
| R.PCA | Q (mL/s) | 1.18 | 1.10 ↓ | **1.14** |
| | P (mmHg) | 96.15 | 89.46 ↓ | 88.38 ↓ |
| | $r_a$ (cm) | 0.0076 | 0.0076 | 0.0077 ↑ |
| R.PCA | Q (mL/s) | 1.18 | 1.12 ↓ | **1.15** |
| | P (mmHg) | 96.14 | 91.01 ↓ | 90.02 ↓ |
| | $r_a$ (cm) | 0.0076 | 0.0076 | 0.0076 ↑ |

Moreover, incorporating an oxygen transport model in this study, which simulates oxygen transport from the heart to the brain, revealed a decrease in $PO_2$ of the R.ACA and R.MCA by approximately 12-14 mmHg compared to the normal state. This finding, previously overlooked, underscores the underestimation of CA effect in prior research and highlights the importance of considering oxygen transport dynamics in such analyses.

3.2. Validation in the clinical dataset

3.2.1 Clinical measurements for clinical dataset

To evaluate the clinical application of our proposed model, the clinical dataset was built including health subject as well as patients. One health subject and 20 patients with the history of ischemic stroke were recruited in this research. Written informed consent was obtained from all patients, and the procedure was approved by the ethics committee of The Eighth Affiliated Hospital of Sun Yat-sen University (Registration No. 2022-075-02). The baseline information was shown in Table 2. However, merely conducting personalized modeling for stroke patients cannot verify the accuracy of our model. The reason is that the PDEs-ODE model describes the dynamic process of CA's regulation of blood flow. Therefore, we introduced a treatment approach to cause changes in the hemodynamics of the patients and verify the accuracy of the model. This treatment method is enhanced external counter-pulsation (EECP). EECP is a non-invasive, guideline-recommended treatment and rehabilitation approach for ischemic diseases like coronary artery disease and stroke. Its therapeutic benefits primarily arise from the improved blood circulation achieved through the sequential mechanical compression of three pairs of pneumatic cuffs, on the lower legs, thighs and hips [29][30], synchronized with the electrocardiogram. EECP improves blood return to the heart and blood perfusion to the upper body. The clinical measurements for 21 people were conducted under two states, rest state and EECP state (30kPa, 300ms duration of the cuffs pressurization).

Table 2(a): The basic information of this subjects

| Patients | Value | Mean ± SD |
|---|---|---|
| Height (cm) | [155.00, 178.00] | 167.00 ± 7.35 |
| Weight (Kg) | [56.00, 80.00] | 67.82 ± 8.26 |
| Age (years) | [26.00, 72.00] | 53.00 ± 13.36 |
| MAP (mmHg) | [88.67, 155.00] | 116.67 ± 17.80 |
| Heart rate (beats/min) | [63.00, 100.00] | 81 ± 9.7 |

First, all participants received either computed tomography angiography (CTA) or magnetic resonance angiography (MRA) to obtain patient-specific Circle of Willis (CoW) structures as shown in Figure 4 (a). For all participants, cerebral blood flow velocity measurements in two physiological states: resting and during EECP at 30 kPa. The blood flow velocity in cerebral peripheral vessels—specifically those identified as stenotic on CTA or MRA, such as the right and left middle cerebral arteries (R.MCA and L.MCA)—was evaluated using Transcranial Doppler (TCD) under both resting and EECP states as shown in Figure 4 (b).

3.2.2 The performances of PDEs-ODE model on clinical dataset

To evaluate the effectiveness of our PDEs-ODE model, we need to build patient-specific models for each participant. Patient-specific models were constructed based on individual imaging data (including TCD, CTA/MRA), that means this personalization process was conducted entirely using measurements obtained under the resting condition, without incorporating TCD data from the EECP state. Once the personalized model was established, the EECP intervention (30 kPa, 300ms) was applied to the model to simulate the resulting changes in cerebral blood flow. The simulation results under the EECP condition were then compared with clinical measurements to evaluate the validity and clinical relevance of the proposed PDEs-ODE model.

In the building of patient-specific models, patient-specific Circle of Willis (CoW) geometries were segmented from MRA/CTA images using ITK-SNAP by a skilled technician. Subsequently, vessel centerlines and corresponding lumen diameter profiles were extracted and used as inputs for the PDEs-ODE hemodynamic model. These individualized CoW structures were incorporated into the PDE-ODE model by replacing the corresponding components (as illustrated in the middle section of Figure 4). Then, we used the GetData software to plot points for the TCD data, and then obtained the blood flow information within one cycle through linear interpolation[28]. After the data was completed gathered, we need a optimization method to obtain patient-specific models. The differences between the PDEs-ODE model with measured blood flow information in the resting state were set as the optimization objects [12]. The optimization method used in this research was the simulated annealing, which has been proven to have advantages in building personalized models [12][28]. During the building of patient-specific model, key hemodynamic parameters, such as MAP, cardiac output amplitude, and heart rate, vessel length and radius, were adjusted to align the simulation results with clinical measurements in the resting state. Until the differences were narrowed down to an acceptable range ( means patient-specific models were obtained), the adjustment of hemodynamic parameters was not stopped. Before the patient-specific models were obtained, they were coupled with EECP model proposed in our previous work [28].

As a representative, comparisons between simulation results and clinical measurements for one patient in both two states were shown in Figure 5. It can be found our patient-specific PDEs-ODE model could describe the hemodynamics features of patients under the effect of CA. Furthermore, the overall comparison across 20 patients were presented in Table 3(a), the one health subject was presented in Table 3(b), using Pearson's correlation coefficient (r) and mean relative error (MRE) [28]. Due to the different locations of the stenosis in each patient, the measurement positions vary. Therefore, the statistics in Table 3 are only related to the stenotic blood vessels.

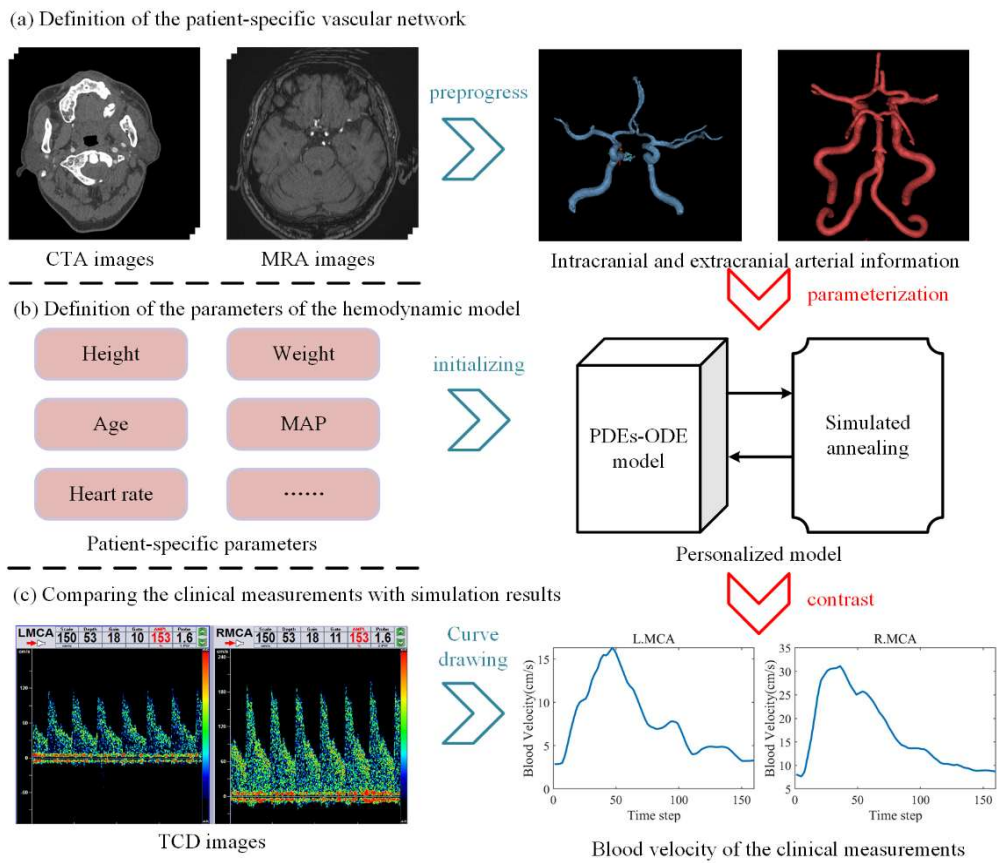

Figure 4. The progress of building the personalized PDEs-ODE model based on CTA/MRA images and TCD images.

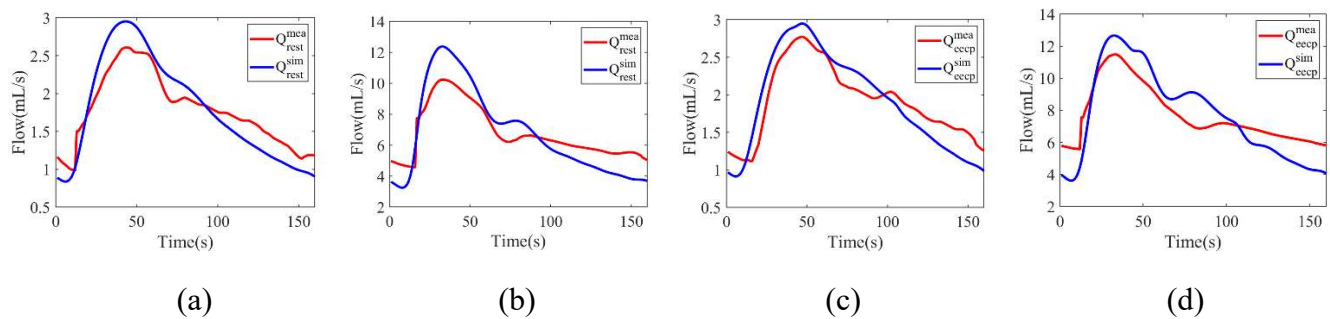

(a)          (b)          (c)          (d)

Figure 5. The comparisons of the personalized results with our PDE-ODEs model and the clinical measurements for one patient in both two states. (a) R.MCA flow in the rest state, (b) L.MCA flow in the rest state, (c) R.MCA flow in the EECP state, (d) L.MCA flow in the EECP state. The blue line represents simulation results of the PDEs-ODE model, and the red line represents TCD measurement results.

Table 3(a): The differences between the clinical measurements and simulation results from patient-specific models for total 20 patients (Two components were measured for each subject, such as the left and right

MCA or the left and right PCA)

|  |  | L.ACA | R.ACA | L.MCA | R.MCA | L.PCA | R.PCA |
|---|---|---|---|---|---|---|---|
|  | Numbers | 2 | 2 | 16 | 16 | 2 | 2 |
| Rest state | $r$ | 0.94 | 0.91 | 0.91 | 0.92 | 0.88 | 0.89 |
|  | MRE | 0.12 | 0.14 | 0.11 | 0.13 | 0.15 | 0.12 |
| EECP state | $r$ | 0.81 | 0.83 | 0.86 | 0.84 | 0.82 | 0.83 |
|  | MRE | 0.17 | 0.16 | 0.15 | 0.17 | 0.24 | 0.18 |

Table 3(b): The differences between the clinical measurements and simulation results from patient-specific models for one health subject in the L.MCA and R.MCA.

|  |  | R.MCA | L.MCA |
|---|---|---|---|
| Rest state | $r$ | 0.92 | 0.94 |
|  | MRE | 0.11 | 0.07 |
| EECP state | $r$ | 0.90 | 0.89 |
|  | MRE | 0.02 | 0.12 |

Table 3(c): The statistic of the measured (white background) and simulated (block background) flow in resting state and EECP state, the unit is mL/s.

|  | Part | Resting(Mean ± SD) | Resting(Mean ± SD) | EECP(Mean ±SD) | EECP(Mean ±SD) |
|---|---|---|---|---|---|
| Patients | R.ACA | 2.19 ± 0.94 | 2.59 ± 1.06 | 2.71 ± 1.06 | 2.95 ± 1.16 |
|  | L.ACA | 2.54 ± 0.20 | 2.88 ± 0.37 | 2.97 ±0.40 | 3.13 ± 0.46 |
|  | R.MCA | 4.24 ± 1.98 | 4.66 ± 2.08 | 4.77 ± 2.12 | 5.00 ± 2.01 |
|  | L.MCA | 3.79 ± 1.65 | 4.17 ± 1.70 | 4.27 ± 1.72 | 4.47 ± 1.78 |
|  | R.PCA | 2.41 ± 0.52 | 2.78 ±0.63 | 2.87 ± 0.60 | 3.04 ±0.71 |
|  | L.PCA | 2.54 ± 0.32 | 2.91 ± 0.49 | 3.18 ±0.47 | 3.00 ±0.43 |
| Health subject | R.MCA | 4.28 | 4.76 | 4.40 | 4.48 |
|  | L.MCA | 4.68 | 5.01 | 4.62 | 5.17 |

Overall, the performances of our PDEs-ODE model in both virtual and clinical dataset prove that it can reliably simulate complex hemodynamics response. Importantly, the hemodynamics changes under effect of EECP can be simulated by our PDEs-ODE model has demonstrated the clinical value of this model.

## 4.Discussion

The goal of our research is to develop a mathematical model to simulate the blood flow under the effect of

CA. According to the widely accepted myogenic hypothesis, CA is mediated by the smooth muscle of cerebral peripheral vessels, which responds to changes in $P$, $PO_2$, and $PCO_2$ levels under ischemic conditions. This mechanism adjusts cerebral peripheral vessel diameter to work. From the perspective of biomechanics, the regulation of the inner diameter of peripheral blood vessels means a decrease in resistance, which makes the blood flow into the vessels easier. This approach can, to some extent, balance the increased vascular resistance caused by vascular stenosis, then maintain stable cerebral blood flow.

Therefore, we developed the PDEs-ODE model to simulate the progress and effect of CA. In the PDEs-ODE model, the distributions of blood flow, pressure, partial pressure of oxygen and partial pressure of carbon dioxide in the human arterial vascular tree were completely completed with the default peripheral vascular inner diameter. These physical distributions of each element were fed into CA module to adjust the radius of cerebral peripheral vessels. After that, the boundary condition of cerebral vessel was reduced, thereby stabilizing the blood flow of the stenosis vessel. To verify its validity, we conducted evaluations on two datasets.

In the virtual dataset, the stenosis models were integrated with the 0D-1D model. It was evident that neglecting the CA model's influence resulted in a reduction of over 20% in blood flow due to 90% MCA stenosis, accompanied by a decrease of MAP in MCA. Conversely, when factoring in the CA module, blood flow levels were restored to those indicative of stenosis-free conditions. Severe ICA stenosis (90%) was also considered in this study, resulting in decreased blood flow in ipsilateral ACA, MCA, and PCA without the influence of CA. After adding the influence of CA, CBF was significantly increased, but the difference from the baseline value remained, which was due to the stenosis of ICA resulting in a substantial reduction of P aO2 and MAP, which led to a weakening of the regulatory ability of CA. These findings imply that patients with ICA stenosis may be at a higher risk of ischemic injury even when global CBF appears acceptable, due to underlying oxygen deficits and reduced CA reserve. Besides that, the simulation tool demonstrates its value in visualizing potential collateral pathways and quantifying CA function on a patient-specific basis. This offers new opportunities for personalized risk assessment and treatment planning. Compared to traditional flow-based models, the integration of oxygen transport and autoregulation modules significantly enhances the physiological relevance and potential clinical applicability of our proposed modeling framework. In the virtual dataset, the validity of the model has been preliminarily verified, but the clinical value of the model still needs to be verified in the actual data.

In the clinical dataset, many patient-specific models were established by adjusting hemodynamic parameters to reduce the differences of $Q_{sim}(t)$ and $Q_{mea}(t)$ in the resting state. As shown in Table 3 and Figure 5 (a)-(d), the personalized PDEs-ODE model had shown great agreement with clinical measurements. But only the results in the rest state cannot prove the effect of PDEs-ODE model, it may be benefited from the powerful personalized method proposed in our previous work. Therefore, we incorporated the influence of EECP into the model (without changing other parameters) to verify whether our PDEs-ODE can consider the effect of CA. As a result, although small discrepancies were noted during the reflected wave period—primarily due to an overestimation of pulse wave velocity in our method—these differences are relatively minor and do not undermine the overall accuracy of the model. This indicates that while there is room for refinement, the model's ability to replicate patient-specific hemodynamics with such precision supports its potential use in personalized medicine. Ultimately, this tool could assist clinicians in understanding cerebral hemodynamics, treatment planning, and prognostication, particularly for complex cerebrovascular conditions where personalized blood flow analysis is crucial.

Although our model has achieved satisfactory results on both two datasets, several limitations still exist: (1) During clinical measurement, it was difficult to obtain the actual distribution of various elements driving CA model (such as $PO_2$, $PCO_2$) in blood vessels. So, the details of PDEs-ODE model were not completely verified. (2) The limited sample size in this study primarily stemmed from concerns regarding potential risks that certain stroke patients might have faced while undergoing the proposed strategy. Overall, our proposed PDEs-ODE model effectively illustrated the impact of CA on maintaining CBF and had been validated across multiple datasets. This enabled the clinical application of this study to generate a digital twin of the hemodynamic environment within subjects' cerebral arteries.

# 5. Conclusions

In this paper, the PDEs-ODE model was constructed by coupling the Navier-Stokes equation, oxygen transport equation, carbon dioxide distribution equation and CA module. The distributions of the arterial pressure, partial pressure of the oxygen and carbon dioxide were calculated and used to drive the CA module. Then the boundary condition of PDEs were adjusted by CA module according to the Poiseuille's law. Until the relative difference in the distribution of physical information in the vascular network compared to the previous cycle is less than 0.5%, the blood flow information under CA regulation is obtained. To verify the accuracy of the model, a virtual dataset containing different stenosis models and a clinical dataset containing 21 subjects were used. In the virtual dataset, the effect of CA on CBF regulation under different conditions was verified. In the clinical dataset, the PDEs-ODE model was proved to can be applied in the clinic after personalization.

**Statements of ethical approval**

**Declarations**

Competing Interests:

No conflicts of interest, financial or otherwise, are declared by all authors.

**Acknowledgements**

**Qi Zhang**: Data progress, Formal analysis, Data curation, Investigation, Writing. **Meng-di Yang** and **Xuan-hao Xu**: Clinical measurements, Data analysis **Li-ling Hao**, and **Shuai Tian**: Supervision, Project administration, Investigation, Funding acquisition, Conceptualization, Resources. Writing.

**Funding**

This work was supported by the Natural Science Foundation of Liaoning Province (No. 2023-MSBA-089),